\documentclass[aps,prl,twocolumn,groupedaddress,bibnotes,showpacs]{revtex4}
\usepackage{graphicx,epsfig,subfigure,afterpage,bm}
\usepackage{afterpage,bm,amsmath}

\newcommand{\be}{\begin{equation}}
\newcommand{\ee}{\end{equation}}
\newcommand{\bea}{\begin{eqnarray}}
\newcommand{\eea}{\end{eqnarray}}
\newcommand{\tauR}{\tau_{\rm R}}
\newcommand{\tauD}{\tau_{\rm D}}
\newcommand{\tw}{t_{\rm w}}
\newcommand{\tf}{t_{\rm f}}

\newcommand{\gdot}{\dot{\gamma}}

\newcommand{\vs}{{\it vs\/\;}}

\newcommand{\etc}{{\it etc.\/\;}}

\newcommand{\bw}{\begin{widetext}}
\newcommand{\ew}{\end{widetext}}

\newcommand{\sig}{\tens{W}}
\newcommand{\wxy}{W_{xy}}

\newcommand{\wyy}{W_{yy}}

\newcommand{\tr}{\mbox{tr}}

\newcommand{\xhat}{\vecv{\hat{x}}}
\newcommand{\yhat}{\vecv{\hat{y}}}
\newcommand{\zhat}{\vecv{\hat{z}}}

\newcommand{\vecv}[1]{\bm{{#1}}}
\newcommand{\tens}[1]{\bm{{#1}}}

\begin{document}

\title{Criteria for shear banding in time-dependent flows of
  complex fluids}

\author{Robyn L. Moorcroft and Suzanne M. Fielding}
\affiliation{Department of
Physics, Durham University, Science Laboratories, South Road,
Durham. DH1 3LE, U.K.}  \date{\today}
\begin{abstract} 
  We study theoretically the onset of shear banding in the three most
  common time-dependent rheological protocols: step stress, finite
  strain ramp (a limit of which gives a step strain), and shear
  startup.  By means of a linear stability analysis we provide a
  fluid-universal criterion for the onset of banding for each
  protocol, which depends only on the shape of the experimentally
  measured time-dependent rheological response function, independent
  of the constitutive law and internal state variables of the
  particular fluid in question.  Our predictions thus have the same
  highly general status, in these time-dependent flows, as the widely
  known criterion for banding in steady state (of negatively sloping
  shear stress {\it vs.}  shear rate). We illustrate them with
  simulations of the rolie-poly model of polymer flows, and the soft
  glassy rheology model of disordered soft solids.
\end{abstract}
\maketitle

Many complex fluids show shear banding~\cite{reviews}, in which an
initially homogeneous sample of fluid separates into layers of
differing viscosity under an applied shear flow.  Examples include
surfactants~\cite{fieldingCates06},
polymers~\cite{PhysRevLett.96.016001}, soft glassy
materials~\cite{ISI:000269862900002,ISI:000175433000002}, and
(possibly) bio-active fluids~\cite{cates2008}. At a fundamental level
shear banding can be viewed as a non-equilibrium, flow-induced phase
transition, or equivalently as a hydrodynamic instability of
viscoelastic origin. In practical terms it drastically alters the
rheology (flow response) of these materials and thus impacts
industrially in plastics, foodstuffs, well-bore fluids, {\it etc}.

In steady state, the criterion for shear banding is
(usually~\cite{noteUsually}) that the underlying constitutive relation
between shear stress $\Sigma$ and shear rate $\gdot$ for homogeneous
flow has negative slope, $d\Sigma/d\gdot<0$. However most practical
flows involve a strong time-dependence, whether perpetually or during
a startup process in which a steady flow is established from an
initial rest-state.  Data in
polymers~\cite{doi:10.1021/ma7027352,hu:275,boukany:73,PhysRevLett.91.198301,ravindranath:957,doi:10.1021/ma9004346,PhysRevLett.97.187801,doi:10.1021/ma802644r},
surfactants~\cite{hu:379,doi:10.1021/ma702527s,huAdd}, soft
glasses~\cite{C1SM05740E,PhysRevLett.104.208301,C1SM05607G,ISI:000280140800011},
and
simulations~\cite{adams:1007,PhysRevLett.102.067801,PhysRevE.76.056106,zhou:591,PhysRevLett.106.055502,ISI:000262976900016,ISI:000285583500029,alexei}
reveals that shear bands often also arise during these time-dependent
flows, and can be sufficiently long lived to represent the ultimate
flow response of the material for practical purposes, even if the
constitutive curve is monotonic, $d\Sigma/d\gdot>0$.

In view of these widespread observations, crucially lacking is any
known criterion for the onset of banding in time-dependent flows.
This Letter provides such criteria, with the same fluid-universal
status as the criterion given above in steady state: independent of
the internal constitutive properties of the particular fluid in
question, and depending only on the shape of the experimentally
measured rheological response function. It does so for each of the
three most common time-dependent experimental protocols: step stress,
finite strain ramp, and shear startup. Our aim is thereby to develop a
unified understanding of experimental observations of time-dependent
shear banding, and to facilitate the design of flow protocols that
optimally enhance or mitigate it as desired.

The criteria are derived via a linear stability analysis performed
within a highly general framework that encompasses most widely used
models for the rheology of polymeric fluids (polymers solutions, melts
and wormlike micelles) and soft glassy materials (foams, dense
emulsions, colloids, {\it etc.}). These general analytical results are
then illustrated by simulations of two specific models: the rolie-poly
(RP) model of polymeric fluids~\cite{ISI:000185523300001}, and the
soft glassy rheology (SGR)
model~\cite{PhysRevLett.78.2020,PhysRevLett.106.055502}.

Throughout we assume incompressible flow, with mass balance
$\nabla\cdot\vecv{v}=0$.  We also assume the flow to be inertialess,
with force balance
$0=\nabla\cdot\tens{\Sigma}=\nabla\cdot(\tens{\sigma}+2\eta\tens{D}-p\tens{I})$.
Here $p$ is the pressure field and $\vecv{v}$ the fluid velocity, with
symmetrised strain rate tensor
$\tens{D}=\tfrac{1}{2}(\tens{K}+\tens{K}^T)$, in which
$\tens{K}_{\alpha\beta}=\partial_\beta v_{\alpha}$.  This generalises
Stokes' equation of creeping flow such that any fluid element carries
a Newtonian stress $2\eta\tens{D}$ of viscosity $\eta$, as in a simple
fluid, and a viscoelastic stress $\tens{\sigma}$ from the internal
mesoscopic substructures in a complex fluid: emulsion droplets,
polymer chains, {\it etc}.

Following standard practice we write $\tens{\sigma}=G\tens{W}$, with
$G$ an elastic modulus and $\tens{W}$ a dimensionless conformation
tensor characterising the deformation of these mesoscopic
substructures. The dynamics of $\tens{W}$ in flow are prescribed by a
rheological constitutive model for the particular fluid in question.
The criteria for shear banding presented below are derived in a
generalized framework~\cite{SM} that includes most commonly used
constitutive models as special cases.  However for pedagogical
purposes we develop our arguments initially within the specific
context of the RP model~\cite{ISI:000185523300001} of polymeric flows,
which has:
\bea
\label{eqn:nRP}
\dot{\sig}+\tens{v}\cdot\nabla\sig&=&\tens{K}\cdot\sig+\sig\cdot\tens{K}^T-\tfrac{1}{\tauD}(\sig-\tens{I})\nonumber\\
&-&\tfrac{2}{\tauR}(1-A)\left[\sig+\beta
  A^{-2\delta}(\sig-\tens{I})\right].  \eea
Here $A=\sqrt{3/\tr{\sig}}$. The terms in $\vecv{v}$ and $\tens{K}$
describe advection by flow, which drives $\sig$ away from undeformed
equilibrium. ($\sig=\tens{I}$ in a well rested fluid.) The remaining
terms model relaxation back to equilibrium: $\tauD$ is the timescale
for a chainlike polymer molecule to escape its entanglements with
other molecules, and $\tauR$ is the (much faster) timescale on which
any stretching of the chain relaxes~\cite{footnote0}.  For convenience
below we often take the non-stretch limit $\tauR/\tauD\to 0$, but
comment on the robustness of our results to this.
Following~\cite{ISI:000185523300001} we set $\delta=-1/2$ throughout.

We consider a sample of fluid sandwiched between parallel plates at
$y=\{0,L\}$, well rested for times $t<0$ then sheared for $t>0$ in one
of the time-dependent protocols defined below: step stress, finite
strain ramp or shear startup. The upper plate moves in the
$\xhat$ direction and the flow is assumed unidirectional, with fluid
velocity $\vecv{v}=v_x(y,t)\xhat$ and shear rate
$\gdot(y,t)=\partial_yv_x$.  Spatial heterogeneity (banding) is
allowed in the flow gradient direction $\yhat$ only, with
translational invariance in $\xhat$, $\zhat$.

The non-stretch RP model then gives, componentwise
\bea
\label{eqn:RPcomp}
\Sigma(t)&=&G\wxy(y,t)+\eta \gdot(y,t),\nonumber\\
\partial_t\wxy(y,t)&=& f(\wxy,\wyy,\gdot),\nonumber\\
\partial_t\wyy(y,t)&=& g(\wxy,\wyy,\gdot),
\eea
with $f=\dot{\gamma} \left[\wyy - \frac{2}{3} (1+\beta)\wxy^2\right]
-\frac{1}{\tauD}\wxy$,
$g=\frac{2}{3}\dot{\gamma}\left[\beta\wxy-(1+\beta)\wxy\wyy \right] -
\frac{1}{\tauD}(\wyy-1)$. Inertialess flow demands uniform total shear
stress: $\Sigma=\Sigma(t)$ only. Our numerics use units in which
$L=1$, $\tauD=1$, $G=1$.

{\it Step stress ---} Consider first a sample subject to a step stress
$\Sigma(t)=\Sigma_0\Theta(t)$ where $\Theta$ is the Heaviside step
function. If the fluid's response to this applied load were one of
homogeneous shear, this would be prescribed by the spatially uniform
but time-dependent solution of (\ref{eqn:RPcomp}):
$\gdot=\gdot_0(t),\tens{W}=\tens{W}_0(t)$.  Differentiating
(\ref{eqn:RPcomp}) shows any such homogeneous state to obey
\bea
\label{eqn:background1}
0&=&G\dot{W}_{0xy}+\eta \ddot{\gamma}_0,\nonumber\\
\ddot{W}_{0xy}&=& \tfrac{\partial f}{\partial \wxy}\,\dot{W}_{0xy}+\tfrac{\partial f}{\partial \wyy}\,\dot{W}_{0yy} +\tfrac{\partial f}{\partial \gdot}\,\ddot{\gamma}_0,\nonumber\\
\ddot{W}_{0yy}&=& \tfrac{\partial g}{\partial \wxy}\,\dot{W}_{0xy}+\tfrac{\partial g}{\partial \wyy}\,\dot{W}_{0yy} +\tfrac{\partial g}{\partial \gdot}\,\ddot{\gamma}_0,
\eea
subject to the initial condition $\gdot_0(0)=\Sigma_0/\eta$,
$\dot{W}_{0xy}=f(0,1,\Sigma_0/\eta)$,
$\dot{W}_{0yy}=g(0,1,\Sigma_0/\eta)$. 

We now examine whether any such state of uniform shear becomes
linearly unstable to the onset of banding at any time during its
evolution. To do so we express the full response to the applied load
as a sum of this underlying homogeneous ``base state'' plus an
(initially) small heterogeneous perturbation: $\gdot(y,t)=\gdot_0(t) +
\sum_n\delta\gdot_n(t)\cos(n\pi y/L)$, $\tens{W}(y,t)=\tens{W}_0(t) +
\sum_n\delta\tens{W}_n(t)\cos(n\pi y/L)$.  Substituting into
(\ref{eqn:RPcomp}) shows that, to first order in
$\delta\gdot_n,\delta\tens{W}_n$, the perturbations obey
\bea
\label{eqn:fluctuation1}
0&=&G\delta{W}_{nxy}+\eta \delta{\gdot_n},\nonumber\\
\dot{\delta W}_{nxy}&=& \tfrac{\partial f}{\partial \wxy}\,\delta{W}_{nxy}+\tfrac{\partial f}{\partial \wyy}\,\delta{W}_{nyy} +\tfrac{\partial f}{\partial \gdot}\,\delta\dot\gamma_n,\nonumber\\
\dot{\delta W}_{nyy}&=& \tfrac{\partial g}{\partial \wxy}\,\delta{W}_{nxy}+\tfrac{\partial g}{\partial \wyy}\,\delta{W}_{nyy} +\tfrac{\partial g}{\partial \gdot}\,\delta\dot\gamma_n.
\eea
These must be solved subject to source terms specifying the seeding of
any heterogeneity, whether due to (i) sample preparation, (ii) slight
flow device curvature, (iii) mechanical or thermal noise.  We consider
(i), using an initial condition $\delta\tens{W}_n(0)=\epsilon_n
\vecv{N}_n$, small $\epsilon_n$, and the entries of $\vecv{N}_n$
drawn from a distribution of mean $0$ and width $1$.

\begin{figure}[t]
  \includegraphics[width=8.5cm]{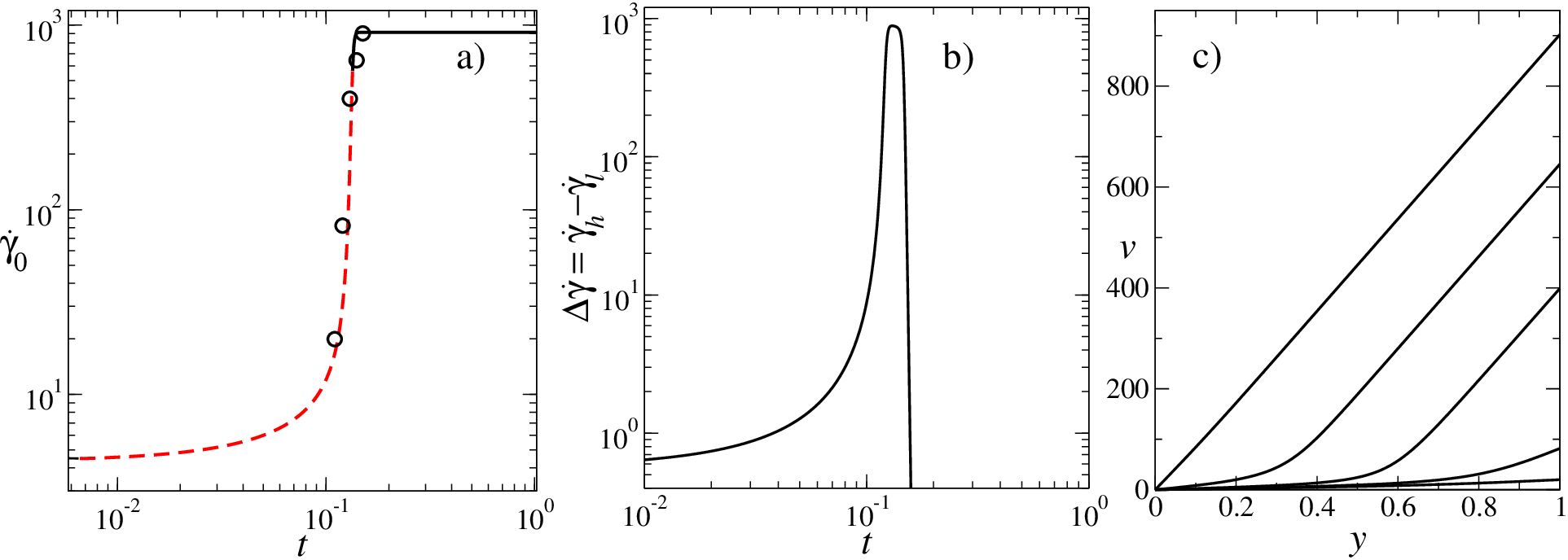}
\vspace{-0.2cm}
\caption{Non-stretch RP model: $\beta=0.8$, $\eta=10^{-4}$,
  $\Sigma_0=0.7$. a) Time derivative of creep curve. Dashed: linearly
  unstable regime.  b) Corresponding degree of banding (difference in
  max and min shear rate across cell). c) Flow profiles at times
  marked by circles in a) for $\epsilon_n=0.1\delta_{n,1}$,
  $l=10^{-2}$. 
}
\label{fig:Jstep}
\end{figure}
%

Eqns. (\ref{eqn:background1}, \ref{eqn:fluctuation1}) together show that the
{\em heterogeneous} fluctuations $\delta\tens{W}_n,\delta\gdot_n$ 
obey the same dynamics as the time derivative of the {\em homogeneous}
base state $\vecv{W}_0,\gdot_0$~\cite{footnote1}.  Shear bands must
therefore develop (growing $|\delta\gdot_n|$) whenever
\be
\label{eqn:Jcriterion}
\frac{d^2\gdot_0}{d t^2}/\frac{d \gdot_0}{d t}>0.
\ee

This criterion is written in terms of the time derivatives of the
creep curve $\gamma_0(t)$ of the underlying base state in our
stability analysis.  How does this $\gamma_0(t)$ relate to the bulk
creep curve $\gamma(t)$ that is measured experimentally by recording
the movement of the rheometer plates?  Clearly, before any banding
develops $\gamma_0(t)=\gamma(t)$ by definition.  Accordingly the onset
of banding out of a state of initially homogeneous creep should happen
once the experimentally measured $\gamma(t)$ likewise obeys
(\ref{eqn:Jcriterion}).

Fig.~\ref{fig:Jstep} shows our numerical results for the non-stretch
RP model, with parameters for which the constitutive curve
$\Sigma(\gdot)$ is monotonic and the steady state homogeneous.
Fig.~\ref{fig:Jstep}a shows a representative time-differentiated creep
curve for homogeneous flow $\gdot_0(t)$. The regime of instability to
banding as predicted by (\ref{eqn:Jcriterion}), where $\gdot_0(t)$
simultaneously shows upward slope and curvature, is shown dashed. A
full nonlinear simulation of the RP model indeed confirms
time-dependent shear banding in this regime (Fig.\ref{fig:Jstep}b,c),
with homogeneous flow recovered in steady state.

How general is this criterion (\ref{eqn:Jcriterion})? Clearly
Eqns.~(\ref{eqn:RPcomp}) -~(\ref{eqn:Jcriterion}) make no assumption
about the functional forms of $f,g$, and so must apply to any
differential constitutive model with $d=2$ dynamical state variables
($W_{xy}$ and $W_{yy}$ above). This is easily extended~\cite{SM} to
arbitrary $d$, to allow for the dynamics of other ({\it e.g.} normal)
stress components, fluidity variables in a soft glass, ordering
tensors in a liquid crystal, \etc Accordingly our criterion
(\ref{eqn:Jcriterion}) should hold for {\em any} constitutive model of
differential form.  Taking $d\to\infty$ extends this to systems with
infinitely many state variables and so, we now also argue, those
governed by {\em integral} constitutive models, of which the SGR model
of disordered soft solids is an example.

\begin{figure}[t]
  \includegraphics[width=8.5cm]{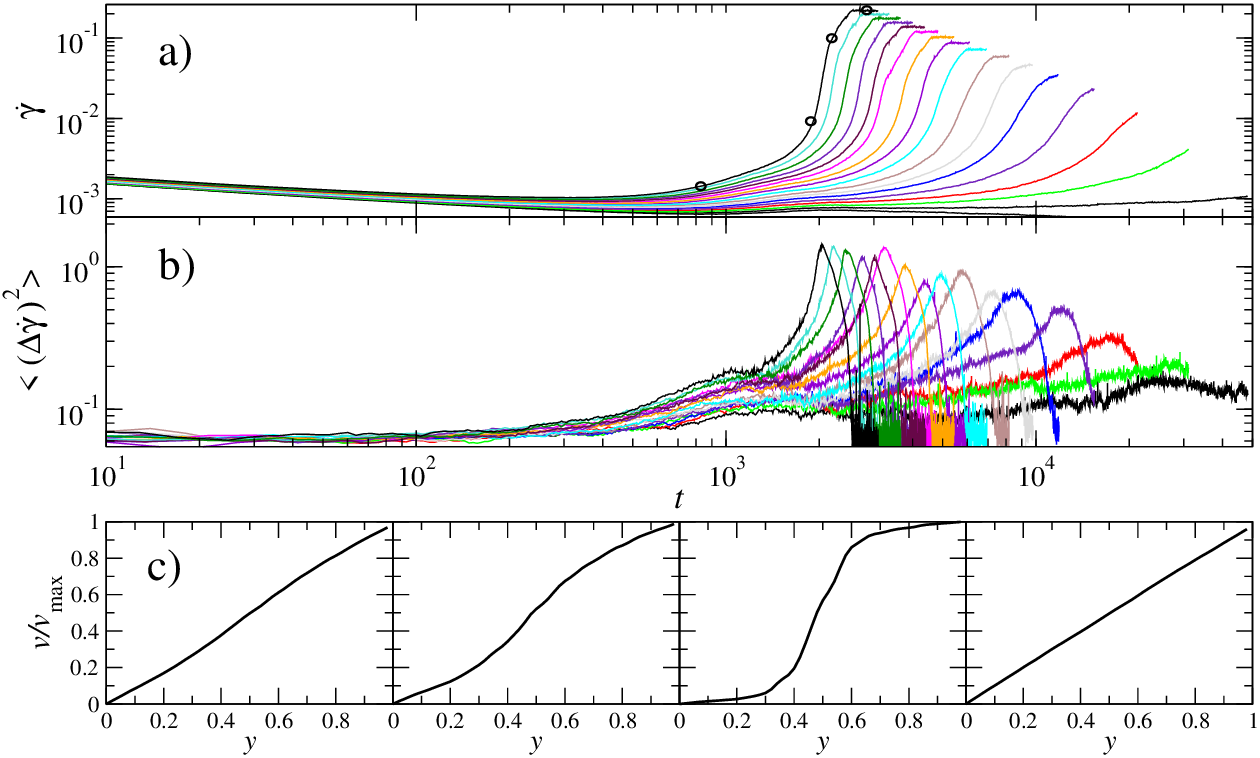}
\vspace{-0.2cm}
\caption{SGR model: a) Differentiated creep curves for stress values
  $\Sigma_0/\Sigma_{\rm y}=1.005,1.010\cdots 1.080$ (curves upwards).
  b) Corresponding degree of banding. c) Normalised velocity profiles
  for the circles in a). $x=0.3, w=0.05,n=50,m=10000$. Initial sample
  age $\tw=10^3\left[1+\epsilon\cos(2\pi y)\right]$, $\epsilon=0.1$.}
\label{fig:SGRcreep}
\end{figure}

Accordingly we now simulate the SGR model~\cite{PhysRevLett.78.2020}
in a form capable of addressing banded
flows~\cite{PhysRevLett.106.055502,SM}.  We focus on its glass phase
$x<1$ where the constitutive curve has a yield stress with monotone
increase beyond: $\Sigma(\gdot)=\Sigma_{\rm y}+c\gdot^{1-x}$.  For an
applied stress just above $\Sigma_{\rm y}$ we see a long regime of
slow creep $\gdot\sim t^{-x}\tw^{x-1}$ , with $\tw$ the sample age
before loading.  See Fig.~\ref{fig:SGRcreep}a. (Experimentally
microgels show $\gdot\sim t^{-2/3}$~\cite{C1SM05607G}, reminiscent of
Andrade creep for plastically deforming crystals~\cite{plastic}.)
This slow creep ends in a transition to a regime of upward slope
$\partial\gdot/\partial t>0$ and curvature $\partial^2\gdot/\partial
t^2 > 0$ in which shear bands form (Fig.~\ref{fig:SGRcreep}b,c),
consistent with (\ref{eqn:Jcriterion}).  Subsequent inflexion to
downward curvature $\partial^2\gdot/\partial t^2<0$ defines a
fluidization time $\tf\sim\tw (\Sigma-\Sigma_y)^{-\alpha}$ with
$\alpha=O(1)$, after which the system recovers homogeneous flow in
steady state.  Microgel experiments~\cite{C1SM05607G} likewise show
$\tf\sim(\Sigma-\Sigma_y)^{-\beta}$ with concentration-dependent
$\beta$.

We therefore finally propose (\ref{eqn:Jcriterion}) as a universal
criterion for shear banding following an imposed step shear stress. It
is consistent with numerous experiments on
polymers~\cite{boukany:73,doi:10.1021/ma702527s,hu:379,hu:1307,PhysRevLett.91.198301,ravindranath:957,huAdd,hu:275}
and soft glassy materials~\cite{C1SM05607G,ISI:000280140800011}.

{\it Finite strain ramp ---} Consider next a well rested sample
subject to a rapid strain ramp $\gamma_0 = \gdot_0 t$ by moving the upper
plate at speed $\gdot_0 L$ for times $0<t<t^*$, after which the strain
is held constant at $\gamma_0^*=\gdot_0 t^*$.  Taking $\gdot_0\to\infty,
t^*\to 0$ at fixed $\gamma_0^*$ gives a true step strain. As above we
shall study this initially in the non-stretch RP model, before
generalising to other materials. 

We start by rewriting~(\ref{eqn:RPcomp}) in a form that emphasizes its
additive loading and relaxation dynamics:
\bea
\label{eqn:RPcomp1}
\Sigma(t)&=&G\wxy(y,t)+\eta \gdot(y,t),\nonumber\\
\partial_t\wxy(y,t)&=& \gdot S(\wxy,\wyy)-\tfrac{1}{\tauD}\wxy,
\eea
with $S=\wyy - \tfrac{2}{3} (1+\beta)\wxy^2$. (The equation for $\wyy$
is not needed here.)  Within this we consider first a state of
idealized homogeneous response to the imposed strain. This will then
form the base state in a stability analysis for the onset of banding
below.  To best approximate a true step strain we focus on a fast ramp
$\gdot \tauD\gg 1$. During any such ramp the base state stress obeys
\be
\label{eqn:before}
\frac{d\Sigma_{0}}{d\gamma_0}=GS(W_{0xy},W_{0yy})\;\;\;\textrm{for}\;\;\; \gdot\tauD\gg 1.
\ee
Post-ramp it relaxes back to equilibrium as $\dot{\Sigma}_{0}=-\Sigma_{0}/\tauD$.

For the fast ramps studied here no banding develops during the ramp
itself.  To investigate whether the sample can remain homogeneous
during its relaxation back to equilibrium, or whether it instead
transiently bands during it, we add initially small heterogeneous
perturbations to the relaxing base state: $\gdot(y,t)=
\sum_n\delta\gdot_n(t)\cos(n\pi y/L)$, $\tens{W}(y,t)=\tens{W}_0(t) +
\sum_n\delta\tens{W}_n(t)\cos(n\pi y/L)$. Substituting these
into~(\ref{eqn:RPcomp1}) shows that, to first order, the perturbations
evolve post-ramp as
\be
\label{eqn:after}
\frac{d\delta\gdot_n}{dt}=-\frac{G}{\eta}S(W_{0xy},W_{0yy})\delta\gdot_n\;\;\;\textrm{for}\;\;\; \eta\ll G\tau. 
\ee

\begin{figure}[t]
  \includegraphics[width=7.25cm]{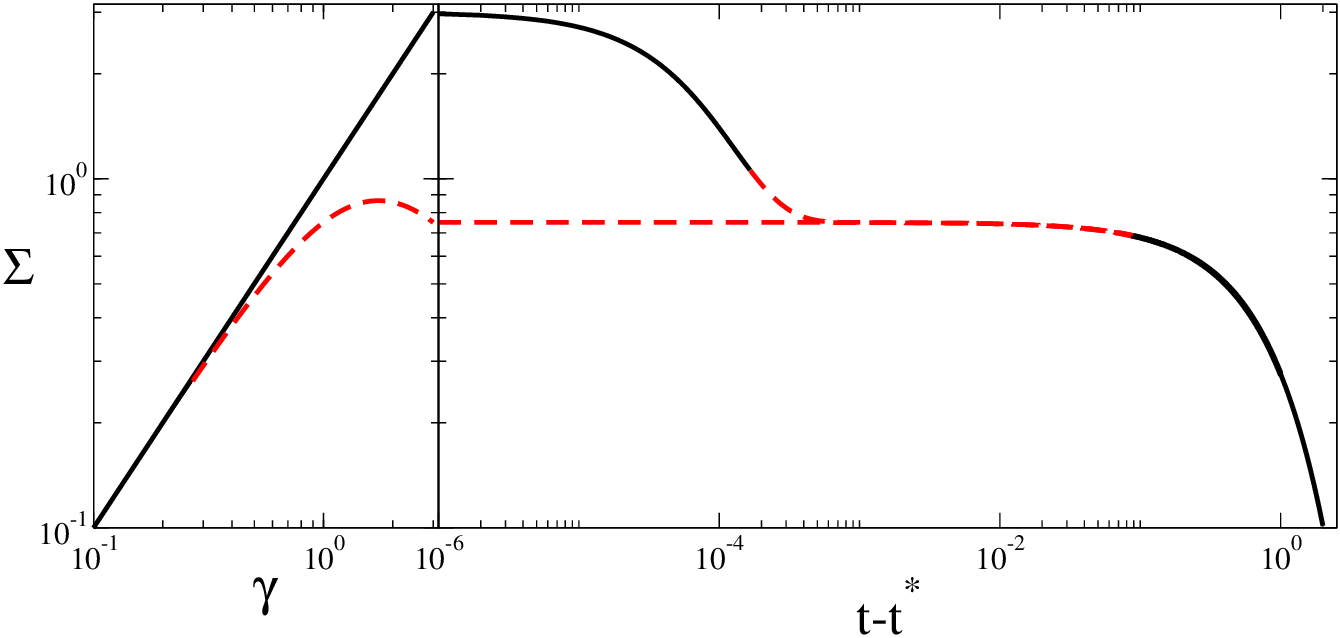}
\vspace{-0.2cm}
\caption{Left: stress {\it vs.} strain for a fast ramp in the
  RP model. $\beta=0.0,\tauR=10^{-4},\eta=10^{-5}$. Right:
  relaxation post-ramp; unstable region dashed. Upper curve:
  appreciable chain stretch, $\tauR\gdot\to\infty$.  Lower:
  negligible stretch, $\tauR\gdot=0.1$.}
\label{fig:Gstep}
\end{figure}

Denoting by $(W_{0xy}^*,W_{0yy}^*)$ the system's state instantaneously
as the ramp ends at time $t^*$, and noting the state to be continuous
at $t^*$, we combine (\ref{eqn:before}) and (\ref{eqn:after}) to show
that the perturbations, immediately post-ramp, obey
\be
\label{eqn:stepResult}
\frac{d\delta\gdot_n}{dt}|_{t=t^{*+}}=\omega\delta\gdot_n\;\;\;\textrm{with}\;\;\;\omega=-\tfrac{1}{\eta}d\Sigma_0/d \gamma_0|_{t=t^{*-}}.
\ee
This shows that shear bands will start developing immediately
following a fast strain ramp if the stress had been decreasing with
strain just prior to the ramp ending
\be
d\Sigma_0/d\gamma_0|_{t=t^{*-}}<0.
\ee
This result accords with early intuition~\cite{marrucci:433}. It can
be shown to hold quite generally~\cite{SM} for all fluids with
additive loading and relaxation dynamics (including the RP model
with chain stretch reinstated).

Numerical results for the RP model support this prediction:
Fig.~\ref{fig:Gstep}.  The lower curve is for a fast ramp in the
non-stretch model.  This has nonlinear loading dynamics, $S=\wyy -
\tfrac{2}{3} (1+\beta)\wxy^2$, so during ramp behaves as a nonlinear
elastic solid with a maximum of stress {\it vs.}  strain.  If the
total applied strain $\gamma^*$ exceeds this, the system is left
unstable to banding immediately post-ramp.  The upper curve shows a
fast ramp in the full model with chain stretch.  This has linear
loading dynamics, $S=\wyy$, and during ramp acts as a linear elastic
solid. Accordingly it is {\rm stable} against banding immediately
afterwards.  However this upper curve reveals further important
polymer physics.  Relaxation of chain stretch on the timescale
$\tau_{\rm R}$ post-ramp restores a state as if no stretch had arisen
in the first place: the upper curve rejoins the lower, both are
unstable to banding and only finally decay on the timescale $\tauD$.
This is consistent with
experiments~\cite{doi:10.1021/ma9004346,PhysRevLett.97.187801,doi:10.1021/ma802644r,fang:939}
and numerics~\cite{PhysRevLett.102.067801,zhou:591} showing that bands
can form either straight after a step strain, or following an
induction period. In extensional equivalent it might also underlie the
physics of delayed
necking~\cite{PhysRevLett.102.138301,ISI:000251451000068}.

The SGR model has linearly increasing stress in a fast ramp so is
stable against banding after it.

\begin{figure}[t]
  \includegraphics[width=8.5cm]{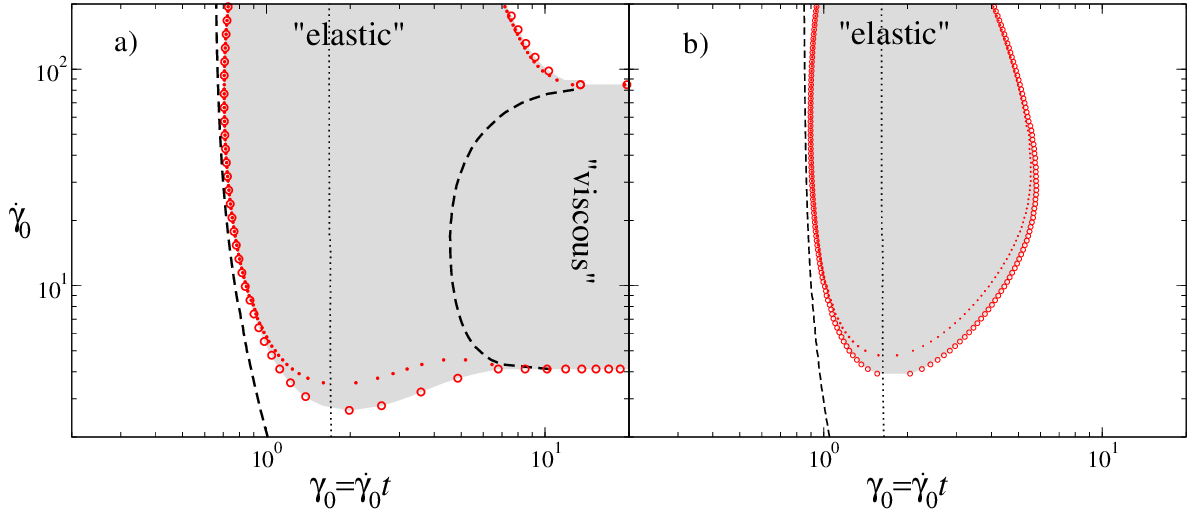}
\vspace{-0.2cm}
\caption{Shear startup in the rolie-poly model. Unstable
  region shaded.  a) Non-monotonic constitutive curve, $\beta=
  0.4,\tauR=0.0,\eta = 10^{-4}$.  Large circles: full onset criterion.
  Right dashed line delimits viscous criterion (\ref{eqn:viscous}).
  Left dashed: elastic criterion (\ref{eqn:elastic2}).  Small circles:
  elastic plus viscous terms (\ref{eqn:elastic2})+(\ref{eqn:viscous}).
  Dotted: stress overshoot $\partial_{\gamma_0}\Sigma_0=0$. b)
  Corresponding figure for monotonic constitutive curve, $\beta=1.0$.}
\label{fig:startupRolie}
\end{figure}

{\it Shear startup --} Consider finally shear applied at constant rate
$\gdot_0$ for all times $t>0$, giving strain $\gamma_0=\gdot_0 t$.
This protocol is discussed here in outline only, with details
elsewhere~\cite{unpublished}.  Our aim is to discover in what regions
of the plane $(\gdot_0,\gamma_0)$ the fluid is unstable to banding
(Fig.~\ref{fig:startupRolie}). Any horizontal slice across this plane
corresponds to the system's evolution in a single startup run at fixed
$\gdot_0$, to steady state in the limit $\gdot_0t=\gamma_0\to \infty$.
A vertical slice at the far right hand side corresponds to the fluid's
steady state properties as a function of $\gdot_0$.

Our calculation~\cite{SM} proceeds as usual by considering a base
state of homogeneous response to this applied shear, then deriving a
criterion~\cite{startupCriterion} for when this becomes unstable to
banding. This contains derivatives of the base state's stress signal
$\Sigma_0(\gamma_0,\gdot_0)$ (which, as discussed above, corresponds
to the experimental signal $\Sigma(\gamma,\gdot)$ at least until
appreciable bands develop).

In a thought experiment in which the flow is artificially constrained
to stay homogeneous until it attains steady state in the limit
$\gamma_0 \to\infty$, this criterion~\cite{startupCriterion} reduces
to the known ``viscous'' instability for steady state bands:
\be
\label{eqn:viscous}
\;\partial_{\gdot_0}\Sigma_0|_{\gamma_0}<0,
\ee
apparent along a vertical slice at the right of
Fig.~\ref{fig:startupRolie}a.

More importantly our criterion~\cite{startupCriterion} also applies to
{\em finite} times $t$ and strains $\gamma_0=\gdot_0 t$. It therefore
predicts at what stage {\em during} startup banding first sets in,
according to the shape of the stress signal as a function of strain
during startup.  Indeed when sheared at a very high rate
$\gdot_0\to\infty$ many materials effectively act as nonlinear elastic
solids, with a stress {\it vs.}  strain curve that attains a unique
limiting function $\Sigma_0(\gamma_0)$, independent of $\gdot_0$. In
any such case our criterion~\cite{startupCriterion} reduces to a
purely ``elastic'' banding instability, onset once
\be
\label{eqn:elastic2}
A\;\partial_{\gamma_0}\Sigma_0|_{\gdot_0} + \gdot_0\partial^2\Sigma_0/\partial\gamma_0^2|_{\gdot_0}<0\;\;\;\textrm{with}\;\;\; A>0.
\ee
The first term, taken alone, predicts onset just after any overshoot
$\partial_{\gamma_0}\Sigma_0=0$ in the stress \vs strain signal. The
second term corrects this, causing onset just before overshoot.  This
is indeed apparent along a horizontal slice at high strain rate in
Fig.~\ref{fig:startupRolie}a.  Eqn.~\ref{eqn:elastic2} holds for any
model with $d=2$ state variables. See~\cite{SM} for $d>2$.

For a fluid with a monotonic constitutive curve,
$\partial_{\gdot_0}\Sigma_0>0$, steady state instability is absent. See
Fig.~\ref{fig:startupRolie}b.  However a patch of elastic-like
instability remains. This shows that shear bands can arise transiently, as
predicted by (\ref{eqn:elastic2}), associated with an overshoot in the
stress startup curve $\Sigma_0(\gamma_0)$, even if absent in steady
state.

Accordingly experimentalists should be alert to the generic tendency
to shear banding in any material that shows an overshoot in stress \vs
strain $\Sigma(\gamma)$ during startup. This may or may not persist to
steady state depending on the slope of the ultimate flow curve
$\Sigma(\gdot)$.  These results are consistent with numerous
experimental~\cite{hu:379,doi:10.1021/ma7027352,boukany:73,huAdd} and
simulation~\cite{PhysRevLett.102.067801,zhou:591,adams:1007,alexei}
studies.

{\it Conclusion --} We have given universal criteria for shear banding
in time-dependent flows of complex fluids. In step stress, banding is
predicted if the creep response curve obeys $(\partial^2\gdot/\partial
t^2)/(\partial\gdot/\partial t)>0$.  In a finite strain ramp, bands
start developing immediately post-ramp if the stress had been
decreasing with strain by the end of the ramp.  In shear startup we
find separate ``viscous'' and ``elastic'' instabilities for a broad
category of fluids that attain a limiting stress startup curve
$\Sigma(\gamma_0)$ in fast flows.  We hope our predictions will help
unify the understanding of widespread data for time-dependent flows,
and stimulate further experiments and simulations of other models
({\it e.g.}~\cite{models}) to test our ideas further.

The authors thank Stephen Agimelen, Mike Cates, Mike Evans, Lisa
Manning, Elliot Marsden, Peter Olmsted, Lewis Smeeton and Peter
Sollich for discussions; and the UK's EPSRC (EP/E5336X/1) for funding.


\begin{thebibliography}{10}
\bibitem{footnote0} In principle diffusive terms $D\nabla^2\tens{W}$
  must also be added to the R.H.S. of Eqn.~\ref{eqn:nRP} to prohibit
  flow heterogeneity at scales smaller than the fluid microstructure
  $l$.  In our linear stability analysis these can be neglected for
  the fluctuations of interest, with wavelength $\gg l$.
\bibitem{footnote1} Due to their different initial conditions they are
  not guaranteed to evolve colinearly.  Numerically, though, we find
  they always do become colinear after a short transient.
\bibitem{startupCriterion} The instability criterion in startup is
  $\partial_{\gdot_0}\Sigma_{0}|_{\gamma_0} -
  \vecv{p}\cdot\tens{M}^{-1}\cdot\left[\partial_{\gamma_0}\vecv{s}|_{\gdot_0}+\gdot_0\,\partial_{\gdot_0}\partial_{\gamma_0}\vecv{s}\right]<0$.
  Here $\vecv{s}$ is a vector comprising all dynamical state
  variables, $\tens{M}$ a matrix formed by linearing the dynamics of
  these state variables, and $\vecv{p}$ a projection vector. See the
  Suppl. Mat. for details.
\end{thebibliography}

\newpage

\end{document}


\title{Supplemental Material: criteria for shear banding in time-dependent flows of complex fluids}

\maketitle

This Supplemental Material comprises two parts. The first details the
two specific rheological constitutive models used to develop our
arguments concerning stability criteria in the main text.  The second
part outlines a theoretical framework that shows these stability
criteria to hold far more widely than these two specific models.

\section{Model definition and components}

\subsection{Rolie-poly model of polymeric fluids}

In the rolie-poly model~\cite{ISI:000185523300001} the dimensionless
conformation tensor evolves according to
\be
\label{eqn:nRP}
\frac{\partial\sig}{\partial t}+\vecv{v}.\nabla\sig=\tens{K}\cdot\sig+\sig\cdot\tens{K}^T-\frac{1}{\tauD}(\sig-\tens{I})-\frac{2}{\tauR}(1-A)\left[\sig+
  \beta A^{-2\delta}(\sig-\tens{I})\right],  
\ee
in which $A=\sqrt{3/T}$ with $T=\tr{\tens{W}}$, and
$K_{\alpha\beta}=\partial_\beta v_\alpha$ where $\vecv{v}$ is the
fluid velocity field. Following~\cite{ISI:000185523300001} we set
$\delta=-1/2$ throughout.  In planar shear flow for which
$\vecv{v}=v_x(y,t)\xhat$, with shear rate $\gdot(y,t)=\partial_yv_x$,
this gives componentwise dynamics
%
\bea
\partial_t \wxy  &=& \dot{\gamma} \wyy \;\; - \frac{1}{\tauD}\wxy \;\;\;\;\;- \frac{2}{\tauR}(1-A)(1+ \beta A)\wxy, \nonumber\\
\partial_t \wyy  &=&\;\;\;\;\;\;\;- \frac{1}{\tauD}(\wyy-1) - \frac{2}{\tauR}(1-A)\left[\wyy+ \beta A(\wyy-1)\right],\nonumber\\
\partial_t T  &=&  2\dot{\gamma}\wxy  - \frac{1}{\tauD}(T-3) - \frac{2}{\tauR}(1-A)\left[T + \beta A(T - 3)\right].\label{eqn:stretch}
\eea
%
These equations fit the generalised framework set out in the next
section below with a $d=3$ dimensional dynamical state vector
$\vecv{s}=(\wxy,\wyy,T)^T$. In the limit of fast chain stretch
relaxation $\tauR\to 0$ they reduce to
%
%
\bea
\partial_t \wxy &=& \dot{\gamma} \left[\wyy - \frac{2}{3} (1+\beta)\wxy^2\right]\;\;\;\;\;\;\; -\frac{1}{\tauD}\wxy,\nonumber\\
\partial_t \wyy &=& \frac{2}{3}\dot{\gamma}\left[\beta\wxy-(1+\beta)\wxy\wyy \right] - \frac{1}{\tauD}(\wyy-1),\label{eqn:nonStretch}
\eea
with constant trace $T=3$, and so with a $d=2$ dimensional dynamical
state vector $\vecv{s}=(\wxy,\wyy)^T$.

In any calculation of shear banded profiles a diffusive term of the
form $D\nabla^2\tens{W}$ must be added to the right hand side of
Eqn.~\ref{eqn:nRP} to prohibit flow heterogeneity arising at scales
smaller than the fluid microstructure $l=\sqrt{D\tauD}$. However in
our linear stability analysis for the onset of banding this term is
negligible for the fluctuations of interest, with wavelength $\gg
l$. Accordingly we omit it in the theoretical analysis that follows
below, and in the main text.

\subsection{The soft glassy rheology (SGR) model}

The SGR model in its original form~\cite{PhysRevLett.78.2020}
considers an ensemble of elements undergoing activated hopping over
local energy barriers $E$, governed by a noise temperature $x$. Each
element is assigned a local strain $l$ which, between hops, obeys
$\dot{l}=\gdot$ where $\gdot$ is the externally applied shear rate.
The characteristic hop time of any element is
\be
\tau=\tau_0\exp\left(\frac{E-\frac{1}{2}kl^2}{x}\right).
\ee
After any hop, an element resets its local strain $l\to 0$ and chooses
a new yield energy $E$ at random from a distribution
$\rho(E)=\exp(-E)$. The evolution of the ensemble's distribution
$P(E,l,t)$ of yield energies and local
strains~\cite{PhysRevLett.78.2020} can be cast exactly in terms of the
corresponding dynamics of infinitely many moments of the probability
distribution~\cite{SMFunpublished}
\be
P_{p,q}=\int_{-\infty}^{\infty}dl\int_0^{\infty}dE\,\frac{l^p}{\tau^q}P(E,l,t),
\ee
to give
\be
\partial_tP_{p,q}=\gdot pP_{p-1,q}+\frac{\gdot q}{x}P_{p+1,q}-P_{p,q+1}+\frac{1}{1+q/x}P_{0,1}\delta_{p,0}\label{eqn:SGR}
\ee
for $p,q=0,1,2\cdots$.  For example $P_{1,0}=\wxy$ (the macroscopic
stress is the average of the local ones) and $P_{0,1}$ is the variable
often described as the material's fluidity. Listing the $P_{p,q}$ in a
state vector $\vecv{s}$, this fits the theoretical framework developed
below with infinite dimensional state vector $\vecv{s}$.

As described in Refs.~\cite{PhysRevLett.106.055502,B812394M}, this
model is rendered capable of addressing banded flows in a numerical
simulation that evolves $j=1\cdots m$ SGR elements on each of
$i=1\cdots n$ streamlines, corresponding to $y=0\cdots 1$ with
periodic boundary conditions. The stress on streamline $i$ is $W_{xyi}
= (1/m)\sum_jl_{ij}$. At any timestep a waiting time Monte Carlo
algorithm chooses stochastically the next element to jump. Supposing
the jump occurs at element $ij$ when its local strain is $l=\ell$,
force balance is then imposed by updating all elements on the same
streamline as $l\to l + \ell/m$.  Diffusive coupling of the dynamics
on neighbouring streamlines is included by further adjusting the
strain of three randomly chosen elements on each adjacent streamline
$i\pm 1$ by $\ell w(-1,+2,-1)$. This spatial diffusivity prevents flow
heterogeneity developing on arbitrarily small lengthscales, as
discussed above in the context of the RP model.

\section{Generalised theoretical framework for stability criterion}

The purpose of this Letter is to provide fluid-universal criteria for
the onset of shear banding in time-dependent flows. In the main text,
for pedagogical simplicity, we developed our arguments in the specific
context of the non-stretch rolie-poly model of polymeric fluids.  In this
section we show that the criteria derived there apply much more
widely: to most commonly used constitutive models for the flow
properties of complex fluids.

As in the main text we consider a sample of fluid sandwiched between
parallel plates at $y=\{0,L\}$, well rested for times $t<0$ then
sheared for $t>0$ in one of the time-dependent protocols defined
below.  The flow is assumed unidirectional, with fluid velocity
$\vecv{v}=v_x(y,t)\xhat$ and shear rate $\gdot(y,t)=\partial_yv_x$.
Spatial heterogeneity (banding) is allowed in the flow gradient
direction $\yhat$ only. The total shear stress $\Sigma$ in any
fluid element is decomposed into a purely Newtonian part of viscosity
$\eta$, and a viscoelastic part due to the polymeric or soft glassy
degrees of freedom: $\Sigma(t)=G\wxy(y,t)+\eta\gdot(y,t)$.  For
the inertialess flows of interest here force balance demands spatially
uniform $\Sigma$, which may however be time-dependent.  The
viscoelastic stress is written as a modulus $G$ times a dimensionless
conformation variable $\wxy$, the dynamics of which are prescribed by
a rheological constitutive equation.

Moving beyond the specific model used in the main text, $\wxy$ is now
taken to obey much more generalised constitutive dynamics. Indeed we
assume only the minimal physical ingredients common to all models for
the rheology of complex fluids, with loading by shear competing with
intrinsic stress relaxation.  In the simplest description
$\partial_t\wxy = \gdot - \tfrac{1}{\tau} \wxy$, though in reality
this shear component also couples to normal
stress~\cite{LarsonComplex} components $\wxx,\wyy$, \etc In soft
glasses it also couples to (at least one) variable characterising the
sample's fluidity~\cite{PhysRevLett.88.175501,PhysRevLett.106.055502}.
In fact the SGR model has an infinite hierarchy of fluidity-like
variables, as described above. In any model of liquid crystal flows it
would couple to the dynamics of the nematic order parameter tensor,
\etc

For any fluid and flow regime of interest, we collect all such
dynamical variables into a state vector $\vecv{s}=(\wxy,\wxx\cdots)^T$
of dimension $d$. Defining the projection vector
$\vecv{p}=(1,0,0\cdots)$, we then have generalised dynamics
\bea
\Sigma(t)&=&G\vecv{p}.\vecv{s}(y,t)+\eta\gdot(y,t),\label{eqn:governing1}\\
\partial_t\vecv{s}(y,t)&=&\vecv{Q}(\vecv{s},\gdot).\label{eqn:governing2}
\eea
(As discussed above diffusive terms $D\nabla^2\vecv{s}$ should also be
included in Eqn.~\ref{eqn:governing2}, but are negligible in the
analysis that follows.)  For specific choices of $\vecv{s}$ and
$\vecv{Q}$, Eqns.~\ref{eqn:governing1} and~\ref{eqn:governing2}
encompass the shear rheology of the rolie-poly model ($d=3$ with chain
stretch, $d=2$ without) and all other constitutive equations of simple
time-differential form. Although the SGR model is conventionally
written in integral form, as discussed above this can be converted to
differential form with an infinite dimensional state vector. Because
our arguments below place no restriction on the dimensionality $d$, we
accordingly argue the criteria that we derive to hold for integral as
well as differential constitutive models.


We now show, for each time-dependent flow protocol in turn, that the
theoretical arguments developed in the main text in the specific
context of the RP model extend in a straightforward way to this much
more general dynamics.

\subsection{Step Stress}

Consider a sample subject at time $t=0$ to a step shear stress:
$\Sigma(t)=\Sigma_{0}\Theta(t)$. Were the material's response to this
loading to be one of homogeneous shear, this would be prescribed by
the spatially uniform but time-dependent solution
$\gdot_0(t),\vecv{s}_0(t)$ of
(\ref{eqn:governing1},~\ref{eqn:governing2}), subject to this applied
$\Sigma(t)$.  Differentiating
(\ref{eqn:governing1},~\ref{eqn:governing2}) it is trivial to show,
for use below, that any such homogeneous state must obey
\be
\label{eqn:background1}
\ddot{\vecv{s}}_0=\left[\tens{M}-G\vecv{q}\vecv{p}/\eta\right]\cdot\dot{\vecv{s}}_0
\ee
with $\tens{M}=\partial_{\vecv{s}}\vecv{Q}|_{\vecv{s}_0,\gdot_0}$,
$\vecv{q}=\partial_{\gdot}\vecv{Q}|_{\vecv{s}_0,\gdot_0}$, subject
to initial conditions
$\dot{\vecv{s}}_0|_{t=0}=\vecv{Q}(\vecv{s}_0|_{t=0},\gdot_0(t=0)=\Sigma_{0}/\eta)$.

To examine whether any such time-evolving state of uniform shear
becomes linearly unstable at any time $t$ to the onset of banding, we
now express the material's full response as a sum of this underlying
homogeneous ``base state'' plus any (initially) small heterogeneous
departure from it: $\gdot(y,t)=\gdot_0(t) +
\sum_n\delta\gdot_n(t)\cos(n\pi y/L)$, $\vecv{s}(y,t)=\vecv{s}_0(t) +
\sum_n\delta\vecv{s}_n(t)\cos(n\pi y/L)$.  Substituting these into
Eqns.~(\ref{eqn:governing1},~\ref{eqn:governing2}), expanding in
powers of $\delta\gdot_n,\delta\vecv{s}_n$, and neglecting terms
beyond first order, shows the heterogeneous fluctuations to obey
\be
\label{eqn:fluctuation1}
\dot{\delta\vecv{s}}_n =\left[\tens{M}-G\vecv{q}\vecv{p}/\eta\right]\cdot\delta\vecv{s}_n.
\ee
This must be solved subject to source terms specifying the seeding of
any heterogeneity, as discussed in the main text.

Straightforward comparison of Eqns.~(\ref{eqn:background1},
\ref{eqn:fluctuation1}) reveals that the {\em heterogeneous}
fluctuations $\delta\vecv{s}_n$ obey the same dynamical equation as
the time derivative $\dot{\vecv{s}}_0$ of the {\em homogeneous} base
state. Due to their different initial conditions, however,
$\dot{\vecv{s}}_0$ and $\delta\vecv{s}_n$ are not guaranteed to evolve
colinearly.  Numerically, though, we find they always do become
colinear after a short transient. Any component $\delta s_{ni}$ of
$\delta\vecv{s}_n$ must then grow whenever its counterpart in
$\vecv{s}_0$ obeys $\ddot{s}_{0i}/\dot{s}_{0i}>0$.  Combined with
Eqn.~\ref{eqn:governing1}, and its linearised counterpart, this means
finally that shear bands must develop (growing $|\delta\gdot_n|$)
following an imposed step stress whenever the differentiated creep
response curve obeys
\be
\label{eqn:Jcriterion}
\frac{d^2\gdot_0}{d t^2}/\frac{d \gdot_0}{d t}>0.
\ee

\subsection{Finite strain ramp}

Consider next a sample subject to a rapid strain ramp $\gamma_0 =
\gdot_0 t$, after which the strain is held constant at
$\gamma_0^*=\gdot_0 t^*$. As above we express the material's response
to this imposed deformation as a ``base'' state of homogeneous shear
across the sample, about which we perform a linear stability analysis
for the dynamics of small shear banding fluctuations.

Assuming decomposition of the viscoelastic dynamics
(\ref{eqn:governing2}) into additive loading and relaxation parts,
$\tens{Q}=\gdot\vecv{S}-\vecv{R}/\tau$, the base state obeys during
ramp
\be
\label{eqn:before}
\frac{d\vecv{s}_0}{d\gamma_0} = \vecv{S}(\vecv{s}_0)-\frac{1}{\gdot_0\tau}\vecv{R}(\vecv{s}_0)\approx \vecv{S}(\vecv{s}_0)\;\;\;\textrm{for}\;\;\;\gdot_o\tau\gg 1.
\ee
This is dominated by loading for the rapid ramps of interest here,
$\gdot_0\tau \gg 1$, which best approximate a true step strain.
Post-ramp the base state relaxes back to equilibrium according to
$\dot{\vecv{s}}_0=-\tfrac{1}{\tau}\vecv{R}(\vecv{s}_0)$.

To investigate whether this relaxation back to equilibrium can indeed
occur in a purely homogeneous way, or if the sample instead
transiently shear bands during it (assuming $t^*$ so short that no
banding arose during the ramp itself), we write $\gdot(y,t)=
\sum_n\delta\gdot_n(t)\cos(n\pi y/L)$, $\vecv{s}(y,t)=\vecv{s}_0(t) +
\sum_n\delta\vecv{s}_n(t)\cos(n\pi y/L)$ and expand to linear order to
find that the heterogeneous perturbations to the homogeneous base
state evolve post-ramp as
\be
\label{eqn:after}
\dot{\delta\vecv{s}}_n=\left[-\frac{G}{\eta}\vecv{S}(\vecv{s}_0)\vecv{p}-\frac{1}{\tau}\partial_{\vecv{s}}\vecv{R}|_{\vecv{s}_0}\right]\cdot\delta\vecv{s}_n\approx -\frac{G}{\eta}\vecv{S}(\vecv{s}_0)\vecv{p}\cdot \delta\vecv{s}_n
\ee
for any fluid in which the Newtonian viscosity is small compared to
the viscoelastic one, $\eta\ll G\tau$.

Denoting by $\vecv{s}_0^*$ the system's state instantaneously as the
ramp ends, Eqn.~(\ref{eqn:before}) shows the evolution of the base
state immediately before the ramp ended to obey
$d\vecv{s}_0/d\gamma_0|_{t=t^{*-}}=\vecv{S}(\vecv{s}_0^*)$.  Because
$\vecv{s}_0$ is continuous at $t^*$, Eqn.~(\ref{eqn:after}) shows the
dynamics of the fluctuations immediately post-ramp to obey
$\dot{\delta\vecv{s}}_n|_{t=t^{*+}}=-\tfrac{G}{\eta}\vecv{S}(\vecv{s}_0^*)\vecv{p}.\delta\vecv{s}_n$.
Combining these, taking the first component, and appealing to the
linearity of force balance (\ref{eqn:governing1}), gives finally
\be
\label{eqn:stepResult}
\frac{d\delta\gdot_n}{dt}|_{t=t^{*+}}=\omega\delta\gdot_n\;\;\;\textrm{with}\;\;\;\omega=-\frac{1}{\eta} \frac{d\Sigma_{0}}{d \gamma_0}|_{t=t^{*-}}.
\ee
This shows that shear bands will start developing immediately
following a fast strain ramp if the stress had been decreasing with
strain just prior to the ramp ending
\be
\frac{d\Sigma_{0}}{d\gamma_0}|_{t=t^{*-}}<0.
\ee
This result holds for a rapid strain ramp in any rheological
constitutive model with additive loading and relaxation viscoelastic
dynamics, and in which the Newtonian contribution to the stress is
small.

\subsection{Shear startup}

Consider finally a shear applied at constant rate $\gdot_0$ for all
times $t>0$, giving strain $\gamma_0=\gdot_0 t$.  As ever we proceed
by considering a base state of uniform flow response to this applied
deformation, then perform a linear stability analysis for the dynamics
of heterogeneous perturbations to this state.

In this shear startup protocol, the most commonly discussed response
function is the stress as a function of strain, for the given strain
rate applied: $\Sigma_{0}(\gamma_0,\gdot_0)$. A familiar thought
experiment then considers a situation in which the flow is
artificially constrained to stay homogeneous until it attains a steady
state of $\gamma_0$-independent
$\Sigma_{0}(\gdot_0),\vecv{s}_0(\gdot_0)$, for $t\to\infty$. In this
limit, with the constraint now removed, the criterion for banding is
well known: $\partial_{\gdot_0}\Sigma_{0}<0$.  Our aim here is to
generalise this result to {\em finite} time $t$, or equivalent strain
$\gamma_0=\gdot_0 t$ during startup to predict at what stage during
the system's transient approach to steady state banding first sets in.

We start by taking
$\partial_{\gdot_0}|_{\gamma_0}\left[(*)/\gdot_0\right]$, where $(*)$
denotes Eqn.~\ref{eqn:governing2}, to find that the homogeneous base
state obeys during startup
\be
\label{eqn:startupBase}
\partial_{\gdot_0}W_{xy0}|_{\gamma_0}=\vecv{p}\cdot\tens{M}^{-1}\cdot\left[\partial_{\gamma_0}
  \vecv{s}_0|_{\gdot_0}+\gdot_0\partial_{\gdot_0}\partial_{\gamma_0}\vecv{s}_0-\vecv{q}\right],
\ee
with $\tens{M},\vecv{q}$ defined as above. It is furthermore
straightforward to show that banding fluctuations about this base
state must obey Eqn.~\ref{eqn:fluctuation1}, as in step stress, the
largest eigenvalue of which crosses zero when
$|-\tfrac{G}{\eta}\vecv{q}\vecv{p}+\tens{M}|=0$.  Combined with
(\ref{eqn:startupBase}) this gives, after some manipulation, the
criterion for the onset of banding during startup as
\be
\label{eqn:criterionFull}
\partial_{\gdot_0}\Sigma_{0}|_{\gamma_0} - G\vecv{p}\cdot\tens{M}^{-1}\cdot\left[\partial_{\gamma_0}\vecv{s_0}|_{\gdot_0}+\gdot_0\,\partial_{\gdot_0}\partial_{\gamma_0}\vecv{s_0}\right]<0.
\ee
(Here and below we neglect small terms involving the Newtonian
viscosity $\eta$.)  In containing not only the ``viscous'' derivative
$\partial_{\gdot_0}\Sigma_{0}$ of the steady state banding criterion
just discussed, but also ``elastic'' derivatives $\partial_{\gamma_0}$
with respect to strain in startup, this onset criterion is valid at
any finite time $t$ and strain $\gamma_0=\gdot_0 t$ in startup.
Clearly in the steady state limit $t\to\infty,\gamma_0\to\infty$ of
any individual startup run at fixed $\gdot_0$ the total accumulated
strain $\gamma_0$ becomes irrelevant and (\ref{eqn:criterionFull})
reduces to the widely known condition for steady state bands:
\be
\label{eqn:viscous}
\partial_{\gdot_0}\Sigma_{0}|_{\gamma_0}<0.
\ee

Conversely, the limit $\gdot_0\to\infty$ at finite $\gamma_0$
corresponds to the transient dynamics of a startup run performed at
very high shear rate. In many cases, the experimentally measured
startup curve of stress {\it vs.} strain attains a unique limiting
function $\Sigma(\gamma_0)$ independent of $\gdot_0$ in this regime:
the material effectively acts as a nonlinear elastic solid with only
``elastic'' derivatives $\partial_{\gamma_0}$ acting in
(\ref{eqn:criterionFull}), giving the simpler onset criterion
\bea 
-\vecv{p}\cdot\tens{M}^{-1}\cdot\partial_{\gamma_0}\vecv{s}|_{\gdot_0}&<&0,\label{eqn:elastic1} \\
-\tr\tens{M}\;\partial_{\gamma_0}\Sigma_0|_{\gdot_0} +
\gdot_0\partial_{\gamma_0}^2\Sigma_0|_{\gdot_0} &<& 0,\;\;\;\;\;\;\textrm{(d=2)}\label{eqn:elastic2}
\eea
with $\tr\tens{M}<0$~\cite{trace}. Here (\ref{eqn:elastic2}) specialises
(\ref{eqn:elastic1}) to the case of just two dynamical degrees of
freedom, {\it e.g.}, the non-stretch RP model. The first term of
(\ref{eqn:elastic2}), taken alone, predicts banding to arise just
after any overshoot in the signal of stress versus strain during
startup, $\partial_{\gamma_0}\Sigma_0<0$; the second corrects this,
causing onset just before overshoot.